**Patient-reported outcomes in the context of the benefit assessment in Germany**

Authors (in alphabetical order):

- Sarah Böhme, Pfizer Deutschland GmbH, Germany
- Christoph Gerlinger, Bayer AG, and Gynecology, Obstetrics and Reproductive Medicine, University Medical School of Saarland, Germany
- Susanne Huschens, Janssen-Cilag GmbH, Germany
- Annett Kucka, Statistical Consultant, Germany
- Niclas Kürschner, Pfizer Deutschland GmbH, Germany
- Friedhelm Leverkus, Pfizer Deutschland GmbH, Germany
- Michael Schlichting, Merck Healthcare KGaA, Germany
- Waldemar Siemens, Roche Pharma AG, Germany
- Kati Sternberg, AbbVie Deutschland GmbH & Co. KG, Germany
- Liping Hofmann-Xu, Bristol-Myers Squibb GmbH & Co. KGaA, Germany

Authors contributed to different sections of the manuscript (authors in alphabetical order):

1) The role of PRO in HTA decision making regarding the benefit assessment in Germany: *Sarah Böhme, Liping Hofmann-Xu, Susanne Huschens, Niclas Kürschner, Michael Schlichting, Waldemar Siemens*
2) Guidance for PRO evaluations: *Susanne Huschens, Niclas Kürschner, Michael Schlichting, Kati Sternberg*
3) PRO Estimand Framework: *Sarah Böhme, Christoph Gerlinger, Friedhelm Leverkus, Michael Schlichting*
4) Perception and requirements for PRO within the German benefit assessment: *Sarah Böhme, Liping Hofman-Xu, Susanne Huschens, Niclas Kürschner, Kati Sternberg*
5) Validity of instruments: *Susanne Huschens, Niclas Kürschner, Kati Sternberg*
6) Response thresholds for assessing clinical relevance of patient-reported outcomes (PRO): *Susanne Huschens, Annett Kucka, Waldemar Siemens, Kati Sternberg*
7) Analyzing PRO endpoints within the framework of the German benefit assessment: *Susanne Huschens, Annett Kucka, Kati Sternberg*
8) Missing PRO data: *Sarah Böhme, Christoph Gerlinger, Friedhelm Leverkus, Michael Schlichting*
9) PRO after treatment discontinuation: *Sarah Böhme, Christoph Gerlinger, Friedhelm Leverkus, Michael Schlichting, Kati Sternberg*

Corresponding Author:
Michael Schlichting, e-mail: michael.schlichting@merckgroup.com

Guest-Editors:
Christoph Gerlinger, Friedhelm Leverkus

# Editorial: Health technology assessment methods for patient reported outcomes in Germany

by

Christoph Gerlinger, Bayer AG, and Gynecology, Obstetrics and Reproductive Medicine, University Medical School of Saarland, Germany

Friedhelm Leverkus, Pfizer Deutschland GmbH, Germany

Regulatory approval is necessary but not sufficient for bringing a Drug to the Patients. A Health Technology Assessment / Reimbursement Process is required in many countries including Germany. Despite several decades of global harmonization in drug development (ICH.org) decisions on the reimbursement of drug are taken using diverging local methodological standards. In the HTA/Reimbursement decisions also societal values and principals play an important role. For example, the utilitarian approach used in England and Wales is deemed unconstitutional in Germany. This makes it rather difficult for international drug development teams to fulfill and understand all countries' needs when designing a drug development program.

The following White Paper by Michael Schlichting and colleagues covers patient-reported outcomes (PRO) and their use for the added benefit assessment in Germany. The paper describes the rules of procedure by the German Federal Joint Committee (Gemeinsamer Bundesausschuss, G-BA) and the General Methods Paper by the Institute for Quality and Efficiency in Health Care (Institut für Qualität und Wirtschaftlichkeit im Gesundheitswesen, IQWiG), which was updated November 5, 2020.

We believe the White Paper by Michael Schlichting and colleagues will help the interested reader to better understand the health technology assessment methods for patient reported outcomes used in Germany.

# Contents

# List of Figures

# List of Tables

# List of Abbreviations

| | |
|---|---|
| AMNOG | Act on the Reform of the Market for Medicinal Products (Arzneimittelmarktneuordnungsgesetz) |
| CHMP | Committee for Medicinal Products for Health Use |
| ClinRO | Clinician Reported Outcome |
| COA | Clinical Outcome Assessments |
| COMET | Core Outcome Measures in Effectiveness Trials |
| CONSORT PRO | Consolidated Standards of Reporting Trials for Patient Reported Outcomes |
| COS | Core Outcome Set |
| COSMIN | Consensus-based Standards for the selection of health Measurement Instruments |
| DIMDI | German Institute for Medical Documentation and Information (Deutsches Institut für Medizinische Dokumentation und Information) |
| EMA | European Medicines Agency |
| EORTC | European Organization for Research and Treatment of Cancer |
| ePRO | Electronic Patient-Reported Outcome |
| EUnetHTA | European network for Health Technology Assessment |
| FDA | US Food and Drug Administration |
| G-BA | Federal Joint Committee (Gemeinsamer Bundesausschuss) |
| GKV-SV | Statutory Health Insurance Funds (Spitzenverband der Gesetzlichen Krankenkassen) |
| HAS | French National Authority for Health (Haute Autorité de Santé) |
| HR | Hazard Ratio |
| HRQoL | Health-Related Quality of Life |
| HTA | Health Technology Assessment |
| IQWiG | Institute for Quality and Efficiency in Health Care (Institut für Qualität und Wirtschaftlichkeit im Gesundheitswesen) |
| ISOQoL | International Society for Quality of Life Research |
| ISPOR | The International Society for Pharmacoeconomics and Outcomes Research |
| ITT | Intention-to-treat |
| LOCF | Last Observation Carried Forward |
| MAR | Missing at Random |
| MCAR | Missing Completely at Random |
| MCID/ MID | Minimal Clinically Important Difference |
| MMRM | Mixed Model Repeated Measures |
| MNAR | Missing Not at Random |
| NICE | National Institute for Health and Care Excellence |

| NSCLC | Non-small Cell Lung Cancer |
| --- | --- |
| ObsRO | Observer-reported Outcome |
| OR | Odds Ratio |
| PerfO | Performance Outcome |
| PRO | Patient-reported Outcome |
| PROM | Patient-reported Outcome Measures |
| QALYs | Quality Adjusted Life Years |
| QoL | Quality of Life |
| RD | Risk Difference |
| RR | Risk Ratio |
| SISAQOL | Setting International Standards in Analyzing Patient-Reported Outcomes and Quality of Life Endpoints Data |
| SMD | Standardized Mean Difference |
| SMPC | Summary of Product Characteristics |
| SPIRIT-PRO | Standard Protocol Items: Recommendations for Interventional Trials for Patient-Reported Outcomes |
| VAS | Visual Analogue Scale |
| vfa | Association of Research-Based Pharmaceutical Companies (Verband Forschender Arzneimittelhersteller) |
| zVT | Appropriate comparative therapy (zweckmäßige Vergleichstherapie) |

# Abstract

Since the 2011 Act on the Reform of the Market for Medicinal Products (Arzneimittelmarktneuordnungsgesetz, AMNOG), benefit dossiers are submitted by pharmaceutical companies to facilitate the Health Technology Assessment (HTA) appraisals in Germany. Following a favorable benefit-risk assessment by regulators, such appraisals should evaluate patient-relevant medical benefit and harm of a new pharmaceutical, a new combination or a new label versus the appropriate comparative therapy (zweckmäßige Vergleichstherapie, zVT), which could be different to those in the pivotal studies. The Institute for Quality and Efficiency in Health Care (Institut für Qualität und Wirtschaftlichkeit im Gesundheitswesen, IQWiG) conducts the added benefit assessment following their General Methods Paper, which was updated November 5, 2020. Alternatively, the Federal Joint Committee (Gemeinsamer Bundesausschuss, G-BA) conducts the benefit assessment (e.g., in case of orphan drug assessments).

The result of the benefit assessment is the foundation for the reimbursement price negotiations with the Statutory Health Insurance Funds (Spitzenverband der Gesetzlichen Krankenkassen, GKV-SV). Appraisal outcomes with "no added benefit" may lead to a reimbursement price based on the cheapest (e.g., generic) appropriate comparative therapy. This could lead to a withdrawal from the market from the perspective of the respective+ pharmaceutical company due to the high economic impact of international price referencing with Germany as a key market. This White Paper is dedicated to patient-reported outcomes (PRO) to highlight their importance for the added benefit assessment. This White Paper is based upon the rules of procedure by G-BA, the methods paper by IQWiG and the analyses of former benefit assessments. We focus on methodological aspects but consider also other relevant requirements and challenges, which are demanded by G-BA and IQWiG. The following topics will be presented and discussed:

1. Role of PRO in HTA decision making exemplary to benefit assessment in Germany
2. Guidances of PRO evaluations
3. PRO Estimand framework
4. Perception and requirements for PRO within the German benefit assessment
5. Validity of instrument
6. Response thresholds for assessing clinical relevance of PRO
7. PRO endpoints / outcome measures / operationalization
8. Missing PRO data
9. PRO after treatment discontinuation

This White Paper aims to provide deeper insights about new requirements concerning PRO evaluations for HTA decision making in Germany, highlight points to consider that should inform global development in terms of study planning and frame the requirements also in the context of global recommendations and guidelines. We also aim to enhance the understanding of the complexity when preparing the benefit dossier and promote further scientific discussions where appropriate.

# 1. The role of PRO in HTA decision making regarding the benefit assessment in Germany

*Authors:     Sarah Böhme, Liping Hofmann-Xu, Susanne Huschens, Niclas Kürschner, Michael Schlichting, Waldemar Siemens*

## 1.1  Patient focused drug development

In recent years, there has been an increased utilization of patient-reported outcomes (PRO) measures and inclusion of the patient voice when evaluating pharmaceuticals or medical technologies. While the inclusion of PRO certainly can have a positive influence on recommendations, varying requirements between different stakeholders can be challenging. The growing use of PRO measures in regulatory and HTA decision making requires robust and scientific sound approaches for PRO data generation, a common understanding what is relevant to patients as well as appropriate statistical analysis methods and their interpretation.

Yet, there are local and global guidances available for regulatory and HTA decision making, that follow different concepts. This remains a challenge for the global patient focused drug development.

This White Paper aims to provide an overview of different perspectives on and requirements for PRO. This paper focusses on the requirements within HTA decision making in Germany. We would like to provide deeper insights on how requirements of the Federal Joint Committee (Gemeinsamer Bundesausschuss, G-BA) and the Institute for Quality and Efficiency in Health Care (Institut für Qualität und Wirtschaftlichkeit im Gesundheitswesen, IQWiG) impact PRO evaluations within the German benefit assessment process. This could guide global analysis planning, analysis conduct and interpretation of PRO outcomes for relative effectiveness assessment complementary to regulatory needs. These requirements differ in parts from the requirements of the regulatory bodies.

## 1.2  Brief introduction to benefit assessment in Germany

In 2011, the Act on the Reform of the Market for Medicinal Products (Arzneimittelmarktneuordnungsgesetz, AMNOG) became effective in Germany. Since then, pharmaceutical companies have to demonstrate an added benefit of a new pharmaceutical, a new combination or a new label versus appropriate comparative therapy (zweckmäßige Vergleichstherapie, zVT) as determined by G-BA. The added benefit is the basis for price negotiations with the National Association of the Statutory Health Insurance Funds (Spitzenverband der Gesetzlichen Krankenkassen, GKV-SV).

The central element is the benefit dossier, which has to be compiled and submitted to G-BA by the local pharmaceutical company. In this benefit dossier, the pharmaceutical company has to demonstrate the added benefit of the new pharmaceutical versus the appropriate comparative therapy. It is of note that the appropriate comparative therapy in the benefit dossier could be different to the comparators used in the pivotal studies. The IQWiG or the G-BA conducts the benefit assessment based on the submitted dossier and provides a recommendation on the added benefit. The conclusive decision on the added benefit rests with the G-BA. The criteria of the benefit assessment are stated in G-BA's code of procedure, in the dossier templates as well as in the General Methods paper of IQWiG (1-3). However, those criteria are often not self-evident. Therefore, a series of articles was authored in 2017 (4) to improve the understanding of the complexity within the scope of the preparation of a benefit dossier and to discuss specific issues.

Since 2017, the AMNOG methodology has advanced rapidly (e.g., analysis of adverse events and PRO). Considering the latest discussions and major changes in the evaluation of PRO in the German benefit assessment, we have identified the need to provide a comprehensive update to the existing article on the topic of health-related quality of life (HRQoL) (5).

## 1.3 What are PRO

This White Paper will refer to the term PRO and consider the following definition in accordance with European Medicines Agency's (EMA) "Appendix 2 to the guideline on the evaluation of anticancer medicinal products in man"(6):

*"A PRO includes any outcome evaluated directly by the patient himself or herself and is based on patient's perception of a disease and its treatment(s). [It reflects] an umbrella term covering both single dimension and multi-dimension measures of symptoms, HRQoL, health status, adherence to treatment and satisfaction with treatment.*

*PRO measures (PROM) are the tools and/or instruments that have been developed to ensure both a valid and reliable measurement of these [PRO]. Like any other clinical outcome assessments such as a rating of a symptom, sign or performance by an observer or trained medical care provider, it is recognised that such data have inherent variability related to the assessor."*

Sometimes, the terms PRO, HRQoL and Quality of Life (QoL) are erroneously used as synonyms. Hence, understanding the differences is important (6).

HRQoL is a multidimensional construct that can be measured in different ways, including direct assessments by the patients. In the following, when referring to HRQoL, we will mainly focus on HRQoL in the context of PRO.

HRQoL "*is a specific type of PRO and is a broad concept which can be defined as the patient's subjective perception of the impact of his/her disease and its treatment(s) on his/her daily life, physical, psychological and social functioning and well-being. The notion of multidimensionality is a key component of the definition of [HRQoL].*" (6).

In the context of health care, HRQoL is used rather than QoL. HRQoL refers to the subjectively perceived state of health or experienced health.

HRQoL is a component of QoL. While QoL focuses on relationships and spirituality, for example, HRQoL focuses on health-related factors. Another difference is that QoL can be measured by the patient or by the doctor or a relative. In contrast, HRQoL is often based on PRO. These are obtained by the patient answering questions in a PRO instrument. All three terms are different and have different meanings; there is an overlap but no substitution.

## 1.4 PRO: different perspectives

Many physicians have access to a highly developed medical technology that can assess physical data, e.g., imaging techniques and laboratory data. However, symptoms like fatigue, headache or anxiety and multidimensional concepts like HRQoL can only be obtained directly from the patient. As patients are the center of any health care system, PRO instruments form a key pillar in health care (7, 8).

In addition, the use of PRO offers opportunities for physicians and health care personnel by improving physicians' satisfaction (e.g., by asking patient-relevant questions), enhancing physician-patient relationships, increasing workflow efficiency (e.g., regular feedback from the patients' perspective), and enabling crucial conversations between physician and patient (9).

Pharmaceutical companies all over the world support the use of PRO in clinical studies as they provide important additional information from the patient's perspective. PRO represent a valid indicator of patient benefit and go beyond the commonly used objective efficacy and safety outcomes.

For the regulatory authorities, the importance of including PRO to assist drawing regulatory conclusions in addition to the conventional efficacy and safety outcomes has been increasingly acknowledged. Both the FDA and EMA have published guidance papers on including PRO in the clinical study (10-12).

From a regulatory perspective (FDA), PRO are embedded in the context of clinical outcome assessments (COA). FDA defines COA as "*a measure that describes or reflects how a patient feels, functions, or survives.*" There are 4 types of COA (10, 11):

- clinician-reported outcome (ClinRO),
- observer-reported outcome (ObsRO),
- patient-reported outcome (PRO), and
- performance outcome (PerfO)

The role of PRO progressed over the last years in drug development. In clinical studies, PRO can be used as primary endpoint and are increasingly used as secondary endpoint impacting PRO labeling, treatment guidance (e.g., S3-Leitlinie) and value generation. Hence, interaction among stakeholders is essential to understand and to reasonably complement clinical trial outcomes considering the patient perspective.

From the HTA perspective, the PRO are an essential component for the benefit assessment. Based on a literature research performed by DIMDI (13) about half of the identified method papers from HTA bodies have discussed the importance of the PRO in the context of benefit assessments and health economic evaluations. There is a difference in the practice across the European countries, though. In France, the transparency commission hardly considers PRO from open label studies. Further, conclusiveness is questioned by the transparency commission if analyses are not adjusted for multiplicity or the minimally important difference (MID) is not defined a priori.

The importance of the PRO in the HTA landscape is particularly underlined in Germany. In the German HTA process (AMNOG), an added benefit of an intervention can only be claimed based on patient-relevant endpoints which include mortality, morbidity, HRQoL and safety (1). PRO can cover the latter two dimensions as they measure the health status (e.g., symptoms), functioning and HRQoL directly from the patients' perspective. Most of all, the measurement of HRQoL based on PRO are crucial for the benefit assessment. For instance, the interpretation of the impact of adverse events (AE) should also be evaluated in terms of how patients perceive associated symptoms and HRQoL as measured by PRO. In some therapeutic settings (e.g., palliative care), an advantage in overall survival (OS) alone might not be adequate to achieve an added benefit, if downsides in AEs are observed. The G-BA requires HRQoL data to judge the overall effect of a new treatment. Not capturing HRQoL in a pivotal trial is regularly criticized by G-BA and has a negative impact on the outcome of the benefit assessment.

## 2. Guidances for PRO evaluations

*Authors:     Susanne Huschens, Niclas Kürschner, Michael Schlichting, Kati Sternberg*

The increasing importance of PROs is also appraised and scientifically discussed within drug regulatory agencies and scientific societies. There are generic guidelines, such as ICH E9 R1 addendum on estimands and sensitivity analysis, that need to be considered for PRO. The most important guidances for PRO are summarized in the following sections.

Table 1: Guidance for PRO evaluations

| Releasing Body | Guidance | Summary |
| --- | --- | --- |
| US Food and Drug Administration (FDA): | Guidance for Industry. Patient-Reported Outcome Measures: Use in Medical Product Development to Support Labeling Claims (14) | e.g., the guidance describes multiple aspects such as:<br><br>• the conceptual framework of a PRO instrument<br>• requirements regarding validity, reliability, ability to detect change<br>• instrument modification<br>• clinical trial design and statistical considerations (e.g., handling missing data, considerations for using multiple endpoints).<br><br>Recently, FDA also published a Draft Guidance for Industry describing the principles for selecting, developing, modifying, and adapting PRO instruments for use in medical device evaluations (15).<br>In addition, FDA published a Patient-Focused Drug Development Guidance Series for Enhancing the Incorporation of the Patient's Voice in Medical Product Development and Regulatory Decision Making to provide further guidance (16). |
| European Medicines Agency (EMA): | Reflection Paper on the Regulatory Guidance for the Use of Health-Related Quality of Life (HRQL) Measures in the Evaluation of Medicinal Products (17) | This paper describes requirements for study design, statistical analysis (e.g., being adapted to address for multiplicity issues and using validated instruments), hypothesis and missing data.<br>Recent developments lead to another EMA reflection paper on the use of PRO and HRQoL measures in oncology studies. This paper takes the increasing importance of PRO into account and describes the framework for drawing regulatory conclusions based on PRO (12). It was followed by an adopted Appendix 2 to the guideline on the evaluation of anticancer medicinal products in man - the use of patient-reported outcome (PRO) measures in oncology studies (6). |

| Releasing Body | Guidance | Summary |
| --- | --- | --- |
| The International Society for Pharmacoeconomics and Outcomes Research (ISPOR) | Measurement Comparability Between Modes of Administration of PROMs (18); Patient Reported Outcomes: Analysis and Interpretation (19); Use of Existing Patient-Reported Outcome (PRO) Instruments and Their Modification (20) | ISPOR offers a variety of initiatives regarding PRO and HRQoL, education and trainings of different topics and offers publications. |
| International Society for Quality of Life Research (ISOQoL) | Implementing Patient-Reported Outcome Measures in Clinical Practice: A Companion Guide to the ISOQOL User's Guide 2018 (21) | ISOQoL has committed to advance the scientific study of HRQoL and other patient-centered outcomes to identify effective interventions, enhance the quality of health care and promote the health of populations. To counter practical challenges for a successful integration of PRO assessments into clinical practice, ISOQoL published a User's Guide to Implementing Patient-Reported Outcomes Assessment in Clinical Practice (the "User's Guide"). The User's Guide is an evidence synthesis that outlines core considerations, and addresses the following 9 questions:<br>1. What are the goals for collecting PRO data in clinical practice and what resources are available? Which key barriers require attention?<br>2. Which groups of patients will be assessed?<br>3. How will the PRO measures be selected?<br>4. How often will the PRO measures be administered?<br>5. How will the PRO measures be administered and scored?<br>6. What tools are available to aid in score interpretation and how will scores requiring clinical follow-up be determined?<br>7. When, where, how, and to whom will results be presented?<br>8. What will be done to respond to issues identified through the PRO assessment?<br>9. How will the value of PRO assessment be evaluated? |
| SISAQOL (Setting International Standards in Analyzing Patient-Reported Outcomes and | International standards for the analysis of quality-of-life and patient-reported outcome endpoints in cancer randomised controlled trials: | SISAQOL Consortium has been convened by the EORTC with the aim to develop recommendations for standardizing the analysis and interpretation of PRO and HRQoL data in cancer trials. The various possibilities of analyzing HRQOL endpoints make it difficult to compare results across different cancer clinical trials. The SISAQOL initiative aims to provide an international recommendation on how to analyze PRO and HRQoL in |

| Releasing Body | Guidance | Summary |
| --- | --- | --- |
| Quality of Life Endpoints Data) Consortium | recommendations of the SISAQOL Consortium (22) | clinical trials. Multidisciplinary working groups were set up on the research objectives, statistical methods and the missing data. |
| Consensus-based Standards for the selection of health Measurement instruments (COSMIN) | The COSMIN checklist for assessing the methodological quality of studies on measurement properties of health status measurement instruments: an international Delphi study (23); COSMIN guideline for systematic reviews of patient-reported outcome measure (24). | Systematic reviews of patient-reported outcome measures are quite complex. The review of a single PRO instrument results in several reviews since for each measurement property of the instrument a review exists. Based on literature research and expert opinions the COSMIN steering committee developed a guideline for systematic reviews. |
| Consolidated Standards of Reporting Trials for Patient-Reported Outcomes (CONSORT PRO) | Reporting of Patient-Reported Outcomes in Randomized Trials. The CONSORT PRO Extension (25). | The CONSORT-PRO is an extension to the CONSORT statement aiming to provide guidance to describe PRO. The extension includes 5 recommended checklist items for RCTs:<br>• identification as a primary or secondary outcome in the abstract<br>• description of the hypothesis of the PRO and relevant domains;<br>• PRO validity and reliability;<br>• statistical approaches for dealing with missing data;<br>• PRO–specific limitations and generalizability of results to other populations and clinical practice. |
| Standard Protocol Items: Recommendations for Interventional Trials for Patient-Reported Outcomes (SPIRIT-PRO) | Guidelines for Inclusion of Patient-Reported Outcomes in Clinical Trial Protocols. The SPIRIT-PRO Extension (26). | The SPIRIT-PRO is an extension to the SPIRIT statement and aims to provide PRO-specific protocol recommendations. The new items focus on<br>• PRO-specific topics in the trial rationale,<br>• objectives,<br>• eligibility criteria,<br>• evaluation of the intervention,<br>• time points for assessment,<br>• PRO instrument selection and measurement properties,<br>• data collection,<br>• translation,<br>• proxy completion,<br>• minimize missing data,<br>• monitoring. |

| Releasing Body | Guidance | Summary |
| --- | --- | --- |
| Core Outcome Measures in Effectiveness Trials (COMET) | Core Outcome Measures in Effectiveness Trials (27). | The COMET Initiative aims to develop a standardized set of outcomes, known as 'core outcome sets' (COS), which should be the minimum to be measured and reported in all clinical trials. COS are also suitable for the use in routine care, clinical audit and research other than randomized trials. The initiative aims to collate and stimulate relevant resources, both applied and methodological, to facilitate exchange of ideas and information, and to foster methodological research in this area. |

To summarize, the ongoing and continuously evolving scientific discussion from different perspectives and players underlines the gaining importance of HRQoL.

# 3. PRO Estimand Framework

*Authors: Sarah Böhme, Christoph Gerlinger, Friedhelm Leverkus, Michael Schlichting*

**Background & rationale**

Clinical study protocols are often less well defined in terms of HRQoL study objectives. For instance, broad objectives are used like "*to evaluate the effect of treatment X on patients' quality of life*" which may lack clarity how to plan analyses.

The estimand framework as described in ICH E9(R1) addendum (28), aims to improve planning, design, analysis and interpretation of clinical trials. Study objectives need to be translated into key research questions, transformed into estimands (= what is to be estimated) of which estimators (= how to estimate) and estimates (= the resulting statistics) can be derived to accurately interpret and conclude about treatment effects also for patient reported outcomes. The estimand is determined by five attributes, i.e.

- treatment
- population
- variable of interest (endpoint)
- population level summary
- intercurrent events

An intercurrent event is defined as an event that occurs after treatment initiation, that affect either the interpretation or preclude observation of the variable associated with the clinical question of interest, e.g., the patient stops treatment or takes additional rescue medication.

How intercurrent events are considered in the analyses may depend on stakeholder perspectives. For instance, similarly to benefit-risk assessments to obtain marketing authorization, there is a need for the HTA authorities to identify the relevant estimands (see the German benefit assessment according to AMNOG, i.e., § 35a SGB V,), that might be different to the regulatory context. There are 5 key strategies to deal with intercurrent events:

- *Treatment policy*: occurrence of intercurrent event is ignored e.g., regardless if patients are on treatment or if new therapy is administered.
- *Composite:* Intercurrent event is a component of variable of interest e.g., if new therapy is considered as part of the PRO outcome.
- *Hypothetical:* scenario in which the intercurrent event would not occur e.g., if patients have not received new therapy.
- *Principal stratum:* target population is subpopulation for which intercurrent event would not occur e.g., patients who receive new therapy.
- *While on treatment:* response to treatment prior to the occurrence of the intercurrent event is of interest e.g., prior to administration of new therapy.

Table 2 provides potential strategies to deal with "treatment discontinuation" as intercurrent event for PRO outcomes:

Table 2: PRO research questions and potential estimand strategies by stakeholders

| Stakeholders | PRO Research Question | Estimand Strategy |
|---|---|---|
| Regulators | How does the treatment impact QoL while the patient takes it as prescribed? | Treatment policy<br>While on treatment<br>Hypothetical |
| HTA (IQWiG) | How will a patient respond in terms of symptoms, functioning, health state given the initial randomized decision to treat? | Treatment policy |
| Patients | What will happen to me if I start this treatment, stop this treatment or if I don´t start treatment at all? What is the effect if I can tolerate the treatment? How long do I have to wait to know whether the treatment is working or not? | Treatment policy<br>While on treatment<br>Hypothetical<br>Principal Stratum |
| Physicians | E.g., for patients with life-threatening conditions, what is the impact of the prescribed treatment, medical care on patient`s quality of life until after the life-saving intervention? | Treatment policy<br>While on treatment<br>Hypothetical |

PRO endpoints, PRO population level summaries, strategies for dealing with intercurrent events including how drop-outs and follow-up times are considered may differ when used in a regulatory context and HTA. Even time-to-event criteria and responder definition could differ according to local HTA requirements answering different research questions.

In this section we focus on relevant PRO endpoints, handling of missing data and long-term follow-up in context of the German benefit assessment and in terms of the estimand framework.

**Requirements by IQWiG**

In general, the treatment policy is the preferred intercurrent event strategy for HTA, e.g., for IQWiG (2). The underlying research question is to evaluate the treatment effect on the variable regardless of the intercurrent event, e.g., switch to subsequent treatment in cancer studies according to protocol, treatment discontinuation due to adverse event.

Post-progression HRQoL can impact HTA decisions, not only in "cost-effectiveness countries" such as UK, Sweden, The Netherlands, but also in Germany. A high potential for bias should be considered in case observation periods vary between groups. Thus, the expectation is that pharmaceutical companies should collect data beyond treatment discontinuation including observations after treatment switching. Further details are provided in Section 9: PRO after treatment discontinuation and Section 8: Missing PRO data.

**Points to consider**

Clinical study protocols are expected to answer multiple research questions for various stakeholders. The estimand framework helps to identify and understand differences and similarities, and potential gaps. HTA decision making is based on estimation of a clinical meaningful effect. Qualitative and quantitative methods should be used to determine PRO thresholds, and adequate endpoints selected to aid meaningful interpretation. Variations of a treatment policy may become useful, e.g., to understand how well treatment works for those who adhere to the prescription guidance and those who do not.

Local HTA guidances may question adequacy of PRO endpoints in terms of instrument validity and what constitutes a meaningful change. For instance, IQWiG may not accept validated or already clinically established MIDs. Universal clinical relevance thresholds were determined for noticeable within-patient changes (see General Methods Paper Version 6.0 (2). More details are provided in Section 6 on response thresholds.

In summary, various estimands and especially intercurrent event strategies may inform evidence synthesis and thus should be considered for treatment specific HTA decision making.

**Discussion and Methodological Recommendations**

More often joint HA and HTA scientific advice meetings should be utilized to consider different perspectives and requirements on the trial design. The primary estimand for HA and HTA may be different. Controlling type I error rate is less relevant in context of G-BA assessment. We believe that sensitivity analysis and/or supplementary analysis utilizing other intercurrent event strategies than treatment policy are useful to support HTA decision making:

- *Treatment policy* strategy cannot be implemented e.g., when considering terminal events such as death (28) – while alive or while on treatment are a more plausible strategy.
- *While on treatment* strategy could reflect certain important patient perspectives, e.g., for contraceptives where the treatment effect should cease immediately after treatment discontinuation.
- *Hypothetical* strategy may provide insights about e.g., if a patient had not received new subsequent anti-cancer therapy.
- *Composite* strategy could combine patient relevant events, for example combining HRQoL deterioration with the intercurrent event death as poor HRQoL outcome.
- *Principal stratum* strategy could address the effect for a relevant subpopulation e.g., patients without intake of pain killers when assessing pain symptom scores.

Such estimands could be carefully considered as a proxy for treatment policy estimand in case PRO data collection is limited but might not necessarily be accepted by IQWiG due to potential bias. Sensitivity analysis may help to address the magnitude of potential bias.

# 4. Perception and requirements for PRO within the German benefit assessment

*Authors:        Sarah Böhme, Liping Hofman-Xu, Susanne Huschens, Niclas Kürschner, Kati Sternberg*

**Background & rationale**

Differences and challenges regarding the PRO measures exist in the context of AMNOG which can be explained by the different perspectives of the G-BA and the regulatory authorities.

The regulatory authorities focus on the hypothesis testing approach, considering primary and secondary endpoints to assess the benefit. FDA and EMA appreciate including PRO in addition to the efficacy and safety endpoints. However, the HTA bodies like G-BA focus on the estimation of the relative effectiveness of the new pharmaceutical versus the appropriate comparative therapy. The AMNOG has, according to German Social Law code book V (SGB V), a much stronger focus on the patient-relevance and therefore also explicitly demands the inclusion of PRO in clinical studies. The hierarchy of endpoints is less important for the added benefit assessment in Germany. Hence, the derivation of the added benefit is based on all patient-relevant endpoints and might results in an added benefit solely based on PRO, which strengthens the role of PRO in German HTA (29-31).

Another essential difference in the perspective is the data collection period of the PRO instruments. Although it is common and well accepted by the regulatory authorities to collect the PRO data until the treatment discontinuation due to e.g., disease progression and lost-to-follow-up of the patients in a clinical trial, the G-BA and IQWiG uses the "treatment policy" and recommends data collection of PRO during the complete study period.

In essence, the comparison of PRO in the AMNOG context is based on a different question than in the regulatory setting. It is more of interest to compare the whole therapeutic schemes rather than the direct treatments.

**Requirements by IQWiG and G-BA**

Specialties in terms of the PRO requirements in the AMNOG context compared to those for regulatory purposes are summarized in Table 3.

Table 3: PRO requirements in the AMNOG and regulatory context

| No. | Topic | AMNOG context | Regulatory context (EMA) |
|---|---|---|---|
| 1. | Presentation of subscales | • Requirements differ depending on the PRO instruments<br>• For e.g., EORTC, data in the Global Health Scores and each subscale should be presented | • Not instrument specific. Data in the summary scores and the subscales should be presented |

| No. | Topic | AMNOG context | Regulatory context (EMA) |
|---|---|---|---|
| | | • For e.g., SF-36, data in the summary scores PCS and MCS should be presented. Data in each subscale or items are optional | |
| 2. | Validity of the PRO instrument | • Only validated PRO instruments will be regarded for benefit assessment<br>• No difference between orphan and non-orphan drugs in terms of the validation criteria | • Generally, the PRO instrument should be validated based on the same criteria that G-BA applies<br>• Niche instruments are not completely rejected, but need to be validated, too |
| 3. | Response thresholds | • An individual response threshold of 15% of the scale range will generally be accepted by IQWiG (see Section 6); G-BA might additionally accept validated/ well-established MID | • A MID is generally accepted, as long as it is well-established in the clinical practice or pre-specified |
| 4. | Return rates | • The G-BA prefers using all randomized/treated patients as denominator to calculate the return rates. Only PRO with at least 70% return rates at each visit will be accepted. | • It is common to include patients who are at risk ("expected patients") in the study as denominator to calculate the return rates<br>• The number of the expected patients varies from time to time and usually decreases during the study |
| 5. | Missing values | • Extent of missing data determines the potential of bias<br>• More than 30% missing PRO data and more than 15%-points of missingness between treatment groups can lead to disregard of the PRO assessment<br>• Statistical method to replace missing data is required | • No threshold for the extent of missingness is defined<br>• The PRO data can still be evaluated using the appropriate replacement methods |
| 6. | Categorization of severity | • Symptoms and HRQoL measured by the PRO instruments can be further categorized to "*severe/serious*" and "*not severe/not serious*"<br>• The extent of the added benefit measured by the PRO is evaluated based on the severity category | • No categorization of severity |

| No. | Topic | AMNOG context | Regulatory context (EMA) |
|---|---|---|---|
| 7. | Significance for continuous PRO endpoints | • Statistical significance in the effect estimate such as the LS-Mean difference alone is not sufficient<br>• Clinical relevance is evaluated by the standardized mean difference (SMD) in terms of Hedges'g, using an irrelevance threshold of 0.2; Hedges'g is required to claim added benefit unless a response threshold (see #3) can be considered; No criteria of the extent of benefit is given for continuous PRO endpoints (i.e. "*non-quantifiable added benefit*"). | • The statistical significance is adequate to evaluate the PRO effect |
| 8. | Follow-up time | • Difference in the follow-up time of the PRO instrument can yield questions regarding the operationalization e.g., in some circumstances, time-to definite deterioration might not be accepted, but rather time-to first deterioration | • Usually, the follow-up time is not the focus of the assessment |

**EQ-5D & QALYs**

Cost utility analyses, for example using EQ-5D to calculate Quality Adjusted Life Years (QALYs), to demonstrate an added benefit are not considered in the German benefit assessment because QALYs do not measure the patients' HRQoL. Instead, it is attempted to approximate the benefit function according to the micro economic theory. These preferences can be used according to the economic welfare theory to determine allocation decisions. Generic measurements of functions, e.g., through EQ-5D in clinical studies, are weighted using tariffs.

Tariffs are an assessment of health conditions which were assessed through a sample of generally healthy insured people (32). Tariffs vary substantially between countries making the results difficult to interpret (33). QALYs are established as the main valuation technique for policy making or for reimbursement decision making in many countries. For example, QALYs are recommended by bodies such as NICE. The QALY concept also remains to be discussed controversial as some experts claim, that the techniques used as a basis for QALY value calculation are flawed. In particular, the underlying assumptions of the multi-attribute utility model do not correspond to behavior patterns observed in a real population (34). This assessment excludes EQ-5D's visual analogue scale (VAS) which is accepted by G-BA as measurement for patient reported morbidity (32, 34).

# 5. Validity of instruments

*Authors:    Susanne Huschens, Niclas Kürschner, Kati Sternberg*

**Background & rationale**

According to G-BA, the analysis of PRO, for example patient reported morbidity and HRQoL demands the same methodological requirements in study design, data analyses and data assessment as for other patient relevant endpoints (35). PRO assessment therefore demands psychometric validated instruments, preferably using a disease specific and a generic instrument (2, 35).

Generic instruments often include core modules and dimensions relevant for most illnesses of interest (e.g., pain, functional impairment), which allow comparisons between interventions or diagnosis groups. Psychometric validation is generally broadly established and referenced. Main challenges include the lack of sensitivity to detect disease specific effects of, for example, the disease of interest.

Disease specific instruments are "custom made" for a specific diagnosis group allowing a more precise distinction between groups and a higher ability to detect changes which could lead to a higher acceptance for patients (36).

The following graph illustrates the required steps of a psychometric validation study (36):

Figure 1: Steps for a psychometric validation study

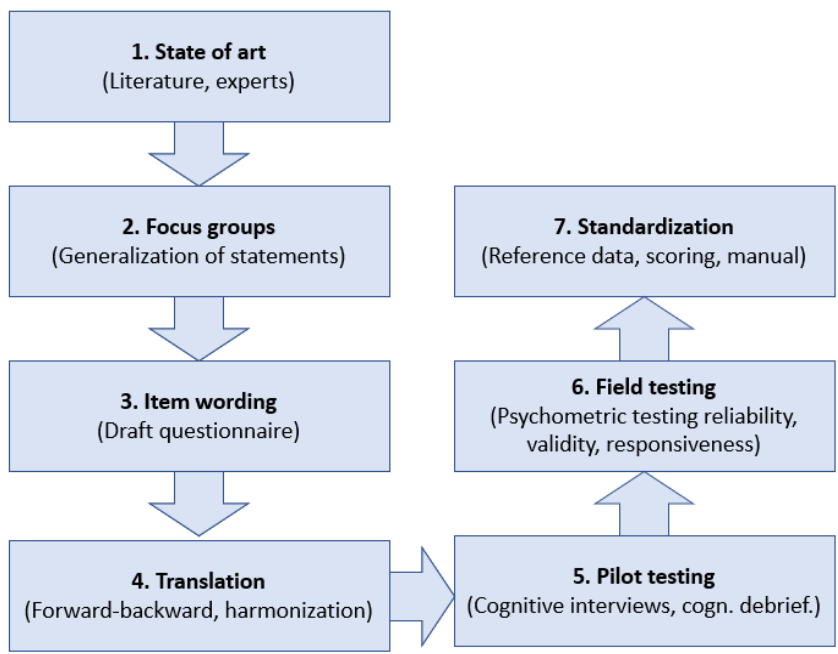

(Adapted according to ((37))

The core aspects of the required psychometric measurement properties are the following (36):

- Reliability: the stability of scores over time when no change is expected in the concept of interest
- Validity: Evidence that the instrument measures the concept of interest
- Ability to detect change: Evidence that a PRO instrument can identify differences in scores over time
- Acceptance & appropriateness: The instrument should have appropriate length and content to allow generation of data suited for the relevant question of interest
- Interpretability: PRO data, especially changes over time should be interpretable in relation to reference data, e.g., from the general population, to allow conclusions based on these quantitative differences

Regulatory bodies, such as EMA and FDA, published guidances in which the requirement for psychometric validation of instruments are described precisely ((14, 15) see also Section 2).

In general, psychometric characteristics of key instruments such as EORTC-QLQ C30 (38), FACT-G (39), SF-36 (40, 41) are well investigated and validated. However, further research might become necessary to confirm validity for a certain study context or a specific therapeutic indication population.

To summarize, these are the requirements by IQWiG and G-BA within the benefit assessment on the validity of PRO instruments:

- The same high standards are applicable as for other patient relevant endpoints
- PRO assessment demands psychometric validated instruments
- Validation studies have to be referenced
- PRO measurement preferably using a disease specific and a generic instrument

For future planning of clinical studies, it is highly recommended to reflect and to assess the validity of potentially implemented PRO instruments at earliest time point possible. If necessary, a validation study should be planned according to the required standards. To allow for more certainty, G-BA early advice regarding study design and planned use of HRQoL instruments is always recommended. This is especially important since G-BA does not provide explicit instructions on how to assure the validity of PRO instruments.

In addition, one of the currently most controversially discussed requirements is the necessity of a validated MID (minimal important difference) for PRO data analyses. Due to its importance, this topic is specifically considered in Section 6.

# 6. Response thresholds for assessing clinical relevance of patient-reported outcomes (PRO)

*Authors:        Susanne Huschens, Annett Kucka, Waldemar Siemens, Kati Sternberg*

**Background & rationale**

Changing the clinical practice by introducing a new drug cannot be based only on statistical significance but has to consider the clinical relevance in terms of the magnitude of improvements the new treatment brings. To evaluate the benefit or harm of a treatment, a specific threshold should be defined to indicate a relevant response. This question is not restricted to PRO but is of special importance for the German benefit assessments due to the strict focus on the patient relevance for all outcomes. The threshold can be viewed from different perspectives. It might be used in the treatment decision of the patients and clinicians, and it is of importance in the approval process and can be the basis for the decision of payers.

The general concept of a patient-individual response threshold must be distinguished from the smallest patient-relevant change in an endpoint resulting in the concept of the "minimal clinically important difference" (MCID; also known as MID). MID can be defined as "the smallest difference in score in the outcome of interest that informed patients or informed proxies perceive as important, either beneficial or harmful, and which would lead the patient or clinician to consider a change in the management" (42). This definition points out the relevance of MID on the individual level (e.g., improvement of pain) as well as on the group level (e.g., revision of treatment guidelines) (43).

Three classes of methods are often used to define the MID: consensus (e.g., Delphi method with an expert panel), distribution-based methods (e.g., 0.5 of the standard deviation), and anchor-based methods (44, 45). While consensus of experts and distribution-based methods per definition fail to incorporate the patient's perspective, anchor-based methods consider the patient's view by examining the relationship between a PRO and a patient-reported anchor (46). For example, the patient's global rating of change can be used as anchor giving a patient the opportunity to rate if she/he felt "*about the same*" "*a little bit better*," or "*quite a bit better*" after receiving a treatment (45).

Devji et al. (2020) point out that, amongst other factors influencing the credibility of MID, the choice of the anchor and the statistical method to estimate the MID are key components in this regard (46). Section 7 provides further methodological considerations with respect to the implementation of responder analysis based on response thresholds or the derivation of the added benefit based on responder analyses.

**Requirements by IQWiG and G-BA**

While a validated or a well-established response threshold in form of a MID is accepted in the regulatory approval process, the IQWiG proposed a new and rather unique concept: All validated or already clinically established MID are questioned and the IQWiG specifies a general response threshold of ≥15 % of the scale range for all questionnaires and all indications. Based on a literature search of systematic reviews, the value of 15% of the scale range was identified as a plausible threshold for a rather small but sufficiently certain noticeable change (see Section 9.3.3 General Methods Paper 6.0 (2)).

It is emphasized by IQWiG that the 15% criterion is not a specific MID but rather a general individual response threshold. It should take into account the empirical variability of MID while being an adequate (but not necessarily the smallest) threshold for a patient-relevant change with the aim to minimize data-driven selection of an MID.

Figure 1 shows 4 scenarios which are discussed in the General Methods 6.0. A prespecified MID of 15% or more of the scale range is always accepted (Scenario 1). Moreover, a responder analysis with the 15 % criterion will be even accepted post hoc, if the prespecified MID is smaller than 15% of the scale range or if no response criterion was prespecified (Scenario 2 and 3). Furthermore, it should be noted that responder analyses are preferred to the analyses of continuous data by the IQWiG (see scenario 4).

Figure 2: Summary of IQWiG's position

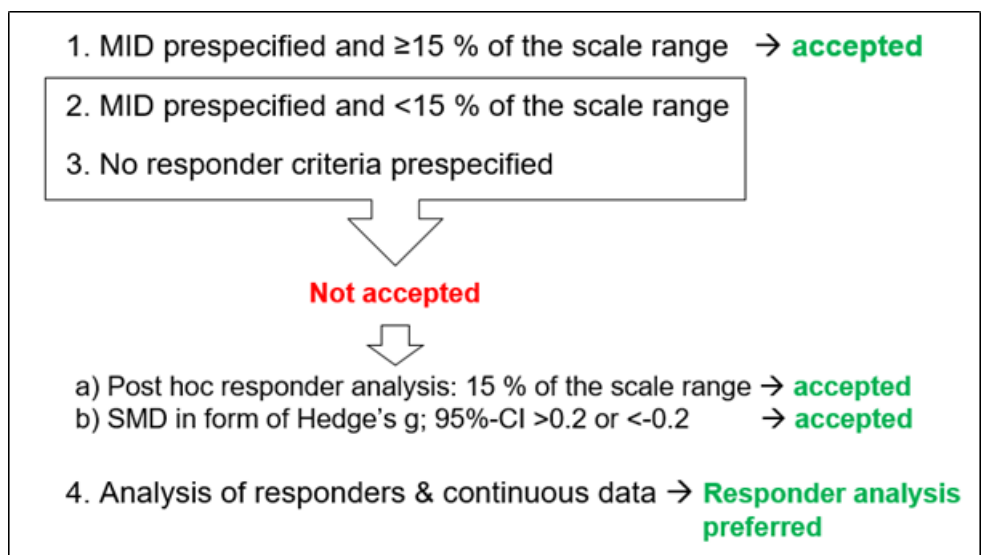

(Adapted according to (2))

**Methodological consideration how to fulfill the requirements for future benefit assessments**

Due to the current methodological considerations on how the requirements for future benefit assessments for MID in responder analyses could look, it is becoming apparent on the basis of the most recent statements by the G-BA that, in addition to analyses with established or prespecified MID, the 15% criterion should also be made available as a response threshold.

The following 3 options exist for submitting responder analyses with an MID. Table 4 presents them and their consequences.

Table 4: Options and consequences in AMNOG of PRO analysis

| Option | Consequences in AMNOG |
|---|---|
| **Continuous analysis of endpoint based on the underlying continuous scale** | **IQWiG:**<br>• used and accepted by IQWiG (if no other responder analysis based on an accepted response threshold available);<br>• IQWiG will propose the derivation of the added benefit based on Hedges'g (see Section 7);<br><br>**G-BA:**<br>• probably accepted;<br>• decision of the added benefit based on Hedges'g; |
| **Analysis of endpoint based on a prespecified and/or clinically established MID** | **IQWiG:**<br>• accepted by IQWiG, if threshold is prespecified and greater than or equal 15% of the scale range or if threshold is not prespecified (post hoc analysis) and equal 15% of the scale range resulting in the proposal of the derivation of the added benefit based on relative risk (RR) or hazard ratio (HR);<br>• not accepted if threshold is smaller than 15% resulting in no added benefit proposed by IQWiG (if not other accepted analysis was delivered);<br><br>**G-BA:**<br>• acceptance unclear;<br>• G-BA might recognize the value of established/validated MID even if the threshold does not meet the IQWiG's criteria;<br>• G-BA might instruct the IQWiG with the re-evaluation of the analysis based on the established/validated MID;<br>• G-BA might request for additional 15% threshold analysis; |
| **Analysis of endpoint based on a response threshold of (at least) 15% of the scale range** | **IQWiG:**<br>• 15% threshold accepted (even post hoc);<br>• Pre-specified threshold >=15% accepted; |

| Option | Consequences in AMNOG |
|---|---|
| | • added benefit derived based on relative risk or hazard ratio (see Section 7 );<br>**G-BA:**<br>• might be accepted;<br>• G-BA might additionally want to focus on the clinically established MID; |

For example, in more recent benefit assessments, the G-BA shows a clear interest in clinically established MID (31, 47), but also, explicitly requires the analysis based on the 15% criterion. In this context, the G-BA bases the decision on the added benefit on the 15 % criterion alone or in addition to the analyses based on established MID (48).

Besides the submission of MID responder analysis, the submission of analyses based on the continuous outcomes of the scales might be an accepted alternative (49). Additionally, it is highly recommended to align on response thresholds in G-BA advisory meetings.

**Discussion and Conclusion**

The 15% criterion, which was published in the General Methods 6.0 by the IQWiG (2), can be regarded as specific requirement in the German benefit assessment. It can be seen as a more rigorous assessment of the established MIDs and a "one-size-fits-all" approach. This approach continues to be discussed at the national level. For example, the Association of Research-Based Pharmaceutical Companies (vfa) demands that the suitability of a response threshold should be examined in each individual case. The actual scientific discussion to define better standards for the evaluation/suitability of response thresholds are also to be supported and a generally accepted catalogue of suitable quality criteria for the assessment of such evaluations should be advanced accordingly. Until then, vfa claims that the usage of all previously accepted assessments based on MID should be continued for reasons of procedural consistency.

On the basis of IQWiG's current General Methods 6.0 and the ongoing discussion on this topic, PRO responder analyses for both established and the 15% criterion should be available for endpoints for which an added benefit is to be derived, in order to be able to sufficiently reflect the requirements for the German benefit assessment. Also, the current requirements for an established MID should also be considered in the emerging MID validation studies (46). How G-BA will finally consider the new IQWiG 15% response threshold in their AMNOG dossier templates is currently under an evaluation procedure at the date of this publication.

# 7. Analyzing PRO endpoints within the framework of the German benefit assessment

*Authors: Susanne Huschens, Annett Kucka, Kati Sternberg*

**Background & rationale**

In the context of the benefit assessment of drugs in Germany, data on PRO such as symptoms, functions and HRQoL outcomes are essential in order to reflect all required categories such as mortality, morbidity, HRQoL and safety in the benefit assessment. Due to various methodological limitations, PRO data is still frequently excluded from assessments by the G-BA and IQWiG.

This may be due to several reasons, for example PRO endpoints representing symptoms, functions and HRQoL are often less well defined in clinical trial protocols than efficacy or safety endpoints. Consequently, PRO endpoints are often analyzed using different methods, which makes it difficult to compare results from different clinical trials and may even lead to different interpretations. This leads to a variety of possible endpoints and outcome measures to determine and operationalize the patient perspective of PRO. Thus, selecting the concept of interest and the correct endpoints to answer the relevant PRO objective remains a challenge.

However, two important guidance documents were published on this topic in 2020:

- ICH E9 (R1) addendum on estimands and sensitivity analysis in clinical trials to the guideline on statistical principles for clinical trials by CHMP; EMA/CHMP/ICH/436221/2017 Committee for Medicinal Products for Human Use, 17 February 2020.
- International standards for the analysis of quality-of-life and patient-reported outcome endpoints in cancer randomized controlled trials: recommendations of the SISAQOL Consortium by EORTC.

These guideline documents are intended to provide a recommendation for defining the appropriate estimator for the clinical question, standardizing the analysis and interpretation of PRO data. This challenge is also reflected in the analysis of PRO endpoints in the German benefit assessment.

**Requirements by IQWiG and G-BA**

The G-BA and IQWiG assess the effects described in the documents presented, taking the uncertainty of results into account. In this assessment, the qualitative and quantitative uncertainty of results within the evidence presented, as well as the size of observed effects and their consistency, are appraised.

From a qualitative perspective the choice and validation of the PRO instrument, if disease specific and generic domains are covered, the frequency and time of measurement and the extent of missing data all affect the assessment of the results. These points are only touched upon in this chapter, for further information please refer to the specific chapters.

Thus, for the assessment of PRO, appropriate instruments are needed that are suitable for use in clinical trials, psychometrically validated, generic and disease-specific (see Section 5). Generic instruments often lack sensitivity to capture disease-specific effects. Hence, the validity of the instrument must also be demonstrated within the disease area. Especially in oncology, there is an increasing demand for a disease-specific questionnaire and MIDs for responder analysis (see Section 6).

The G-BA also required a minimum of 24 weeks treatment and observation of endpoints, especially for chronic diseases. These kinds of requirements should be discussed specifically for the indication in the context of a G-BA advisory meeting. Data collection after treatment discontinuation is also a controversial topic, so it is specifically considered in a separate section within this White Paper (see Section 9).

Another important point in the context of the German benefit assessment is the timing of the analysis of PRO endpoints. Usually, the G-BA prefers a later time point, but also considers potential sources of bias (large amount of missing data, low return rates, small risk set). In contrast, a change in treatment (planned treatment switch or new follow up treatment) might not necessarily be seen as a source of bias (see Section 9) as long as this concept meets the clinical practice or the treatment guideline.

The data cut should be available for all required categories such as mortality, morbidity, HRQoL and safety. This also applies if a data cut was originally planned only for the analysis of specific endpoints. The presentation of the results of individual endpoints of a data cut or of an entire data cut can be dispensed with if no significant additional information is to be expected compared to another data cut (e.g., if the follow-up of an endpoint was already almost complete for the previous data cut or if a data cut is in direct temporal proximity to another data cut).

According to G-BA and IQWiG, the same principles apply to the analysis of PRO as to other endpoints, including the submission of subgroup evaluations. In contrast to regulatory authorities, detailed subgroup analyses for the benefit assessment have to be conducted for all endpoints, not only for the primary endpoint. Even though PROs are often included as exploratory endpoints in the study protocol, additional analyses are required by the G-BA and IQWiG in order to conduct a full assessment. Consequently, subgroup analyses have to be presented for each PRO score that can be e.g., symptom, function or HRQoL question. The number of subgroups depends on the subgroups defined in the study SAP.

**Analyzing PRO endpoints and requirements for the derivation of added benefit by G-BA and IQWiG**

The basic concept, for the assessment by the G-BA and IQWiG, is to derive thresholds for confidence intervals for relative effect measures e.g., hazard ratios depending on effects to be aimed for, which in turn depend on the quality of the target variables and the magnitude categories.

The following scales can be distinguished:

- Binary scales
- Continuous or quasi continuous with responder analyses available in each case (analyses of mean values and standard deviations)
- other (e.g., analyses of nominal data).

In the context of the benefit assessment, an extent of added benefit can be claimed using the categories: major, considerable, minor, non-quantifiable added benefit, no added benefit proven, benefit of the drug to be assessed less than the benefit of the appropriate comparator therapy.

However, it is not always possible to quantify the magnitude of the added benefit for an endpoint, depending on how the PRO endpoint was analyzed. For example, quantification of the effect for the PRO endpoints is possible in the following cases:

*Analysis of binary outcomes*

To determine the magnitude of the effect for binary outcomes, the 2-sided 95% CI for relative risk (RR) will be used. Based on these 2 parameters, the added benefit can be derived. It is recommended that the output tables list the proportions with their 95% CI as well as the measures odds ratio (OR), relative risk (RR), and risk difference (RD), but there are no precise specifications from the G-BA and IQWiG on this with respect to methods.

For binary variables based on continuous endpoints, an accepted response threshold must be used. For further details, see Section 6. Responder analyses based on response thresholds (improvement and/or worsening) are then handled analogously to other binary endpoints.

The IQWiG explicitly requires a time-to-event analysis for binary outcomes with different observation times between treatment arms. This usually applies to oncology studies with increasingly different follow-up times between treatment arms. In time-to-event analyses, PRO data can be evaluated in several ways, e.g., the time-to-deterioration and the time-to-improvement analyses. The results are presented as hazard ratios with 95% CIs and Kaplan-Meier plots, from which added benefit is derived.

Worsening and improvement are defined as an increase or decrease of X points from baseline, according to the accepted response criterion described above (see Section 6). Whether an increase or decrease is considered an improvement or worsening by the patient depends notably on the indication or questionnaire and respective items.

Time-to-event analyses can be performed that consider only the first event or sustained improvement or deterioration. Analyses based on sustained improvement or deterioration take into account more information and may be more meaningful in some situations.

Consideration should be given to the fact that individual subjects may improve as well as deteriorate over the course of the study, so that the change in these individual subjects does not appear to be permanent. In these cases, it may be appropriate to conduct an alternative analysis that accounts for sustained improvement or deterioration.

Also, in the case of fluctuating results with continuous endpoints, analyses can be carried out with a Mixed Model Repeated Measures (MMRM) that takes the entire period into account.

*Analyses of continuous or quasi-continuous outcomes*

In analyses of continuous or quasi-continuous data, effect estimates such as mean treatment difference including confidence interval should be reported. A general statistical measure in the form of standardized mean differences (SMD, in the form of hedges' g) and the associated confidence interval is used to derive the added benefit. For other outcome measures for which no responder evaluations with derivable relative risks (RR) are available, it must be assessed on a case-by-case basis whether relative risks (RR) can be approximated to apply the appropriate thresholds for determining magnitude. Otherwise, the magnitude is to be determined as non-quantifiable.

Specifically, for PRO, the G-BA requires a graphic representation as well as an analysis over the course of the study. Here, the G-BA mentions an analysis that includes all values during the course of the study with explicit reference to MMRM. The pharmaceutical company has to decide, if the graphical overview might be based on mean values or on the response rates based on a discussed threshold at each individual time point. Furthermore, the use of raw values compared to adjusted values in the graphics e.g., from the used statistical model (e.g., MMRM) should be determined. In general, the suitability of an MMRM has to be discussed depending on the specific PRO and study design. Graphical representation and MMRM might be used by the IQWiG to verify the persistence of effects, might support the specific choice of the time point to derive the added benefit and might be used to discuss fluctuating changes in an endpoint. Usually, these analyses are presented additionally to the analysis to derive the added benefit.

*Derivation of added benefit by G-BA and IQWiG*

An added benefit can be quantified using the estimator, RR and HR, and the associated 95% CI. The confidence interval must lie completely below a certain threshold for the extent to be regarded as minor, considerable or major. The critical (decisive) factor is that the upper limit of the confidence interval is smaller than the stated threshold. The thresholds are specified separately for each category. The more serious the event, the bigger the thresholds (i.e., closer to 1). The greater the extent of the added benefit, the lower the thresholds (i.e., further away from 1) for the upper confidence limit for the RR (OR, HR). Table 5 provides the critical values below which the upper confidence limits must lie for the three extent categories of benefit (minor, considerable, major) and types of outcome (2, 50). For e.g., a PRO referring to non-serious symptoms the confidence interval of the RR or HR must be completely below 0.9.

Outcomes can be categorized as severe or non-severe symptoms or HRQoL. The rationale for this operationalization is presented in the methodological approach for determining the extent of added benefit as well as in Skipka et al (50).

Table 5: Thresholds for determining the extent of an effect

|  |  | **Outcome category** | | |
|---|---|---|---|---|
|  |  | All-cause mortality | Serious (or severe) symptoms (or late complications) and adverse events, as well as health-related quality of life [a] | Non-serious (or non-severe) symptoms (or late complications) and adverse events |
| **Extent category** | Major | 0.85 | 0.75 and risk ≥5 % [b] | Not applicable |
|  | Considerable | 0.95 | 0.90 | 0.80 |
|  | Minor | 1.00 | 1.00 | 0.90 |
| a: Precondition (as for all patient-reported outcomes): use of a validated or established instrument, as well as a validated or established response criterion | | | | |
| b: Risk must be at least 5 % for at least 1 of the 2 groups compared. | | | | |
| Literature Reference (2) | | | | |

**Discussion/Interpretation of effect direction**

The testing for a difference between the treatment arms is carried out by comparing "intervention to be assessed" vs. "appropriate comparison therapy". Exceptions are those endpoints that indicate a positive event. In these cases, the comparison between the two treatment arms is carried out in reverse ("appropriate comparison therapy" vs. "intervention to be assessed"). This ensures that an effect <1 (negative events) is an advantage in favor of "intervention to be assessed" and an effect <1 (positive events) is a disadvantage of "appropriate comparison therapy" for all outcomes. According to the IQWiG methods paper, this procedure should be followed to be able to determine the extent of the effect according to Table 5.

**Discussion and Conclusion**

Both the G-BA and IQWiG require data on PRO to be presented as part of the benefit assessment of drugs. This is also confirmed by the ongoing methodological discussions on PRO. The biggest challenge for the pharmaceutical company in the benefit assessment in Germany is to quantify a qualitative effect variable and to estimate its value. In many cases where data from suitable PRO instruments are available for assessment by the G-BA and IQWiG, no additional benefit can be derived due to the different requirements for validation of PRO instruments or methodology.

These limitations may lead to the PRO not being assessed or to a lack of statistical significance or missing patient data due to methodological limitations. In order to address the relevance of PROs in the context of the benefit assessment of medicinal products and the associated requirements, discussions such as those of the ICH E9 (R1) Addendum and the SISAQOL recommendations should be further advanced.

# 8. Missing PRO data

*Authors:     Sarah Böhme, Christoph Gerlinger, Friedhelm Leverkus, Michael Schlichting*

**Background & rationale**

PRO are increasingly used to inform decisions on benefit-risk and relative effectiveness. There is common sense that complete data capture from all patients is essential to limit bias and ensure generalizability for the population of interest. Therefore, every effort should be undertaken by study pharmaceutical companies to fulfil all the requirements of the protocol concerning the collection and management of data (51). Electronic data collection of PRO by the patient is increasingly used to reduce amount of missing data. However, it is unavoidable that data might be missing as patients drop-out due to several reasons, measurements cannot be taken as scheduled due to administrative or personal reasons. In addition, there are unobserved assessments for instance after data cut-offs, after key study objectives are met. It is also important to account for data that is not collected after death. Unobserved assessments may not be necessarily considered as missing data but will affect the definition of the estimand.

The type and proportion of missing data will impact if and how study research questions can be answered. Potential bias to draw conclusions about any treatment effect is influenced by the relationship between missingness, treatment assignment and outcome (52, 53). Little and Rubin defined three missing data mechanisms:

- missing completely at random (MCAR),
- missing at random (MAR) and
- missing not at random (MNAR)

The definition of PRO estimands, in particular the intercurrent event strategies provide a framework for how missing data are handled. Events after first dosing or randomization can result in missing data and thus affect PRO interpretation (e.g., withdrawal, treatment discontinuation, disease progression, non-compliance, administration of rescue medication). The intercurrent event strategy and sensitivity analyses aim to address missingness a-priori to ensure that the estimate is appropriate to answer the research question.

In this part of the paper, we will focus on two types of missing data (54) that are specific for multi-dimensional, multi-item PRO instruments:

- item non-response (responses on some items are missing) and
- unit non-response (the whole questionnaire is missing). Unit non-response can be due to patient drop-out from the study, intermittent missing questionnaires

Missing data issues related to study design considerations such as missing due to variation in follow-up times is handled in the Section 9.

**Requirements by IQWiG and G-BA**

Like regulatory requirements such as ICH E9, HTA is focused to ensure that all randomized patients can be included in the statistical analysis and to avoid and minimize potential for bias also by adopting adequate methods to impute missing values. IQWiG requires a treatment policy estimand i.e., an estimate of the effect for the entire treatment strategy independent of the intercurrent event (e.g., treatment discontinuation, patient drop-out, disease progression, switching of treatment) following an intention-to-treat principle (2). Hence, any data missing for the corresponding main estimate is critical. According to the required treatment policy estimand, data is considered missing even if unobserved according to study protocol. Consequently, the PRO return rates (also called PRO compliance rates) should be calculated using all randomized/treated patients in the denominator.

Missing PRO data limits the acceptance for HTA decision making. A lack of sufficient HRQoL data, for example, can lead to a downgrading of the added benefit. Specifically, there are concrete expectations and thresholds to define what is a substantial number of missing data considered in the benefit assessment in Germany. PRO data are in general not considered when more than 30% of the study population is completely excluded from the analysis. In addition, PRO analysis will also be disregarded in case the rate of excluded study participants between treatment groups exceed 15%-points accordingly (2). Such a finding would promote a missing not at random assumption (MNAR). Furthermore, IQWiG evaluates the potential for bias as introduced by lack of follow-up, e.g., by drop-outs. The magnitude and timing, reasons for drop-outs and particularly differences between treatment groups are to be investigated. Whether the PRO analysis can be considered robust or not may depend on the ability of selected replacement strategies or statistical analysis methods to compensate for bias (2). A detailed description and understanding of the mechanism leading to lost to follow-up is required.

In summary, methods such as imputation of missing values using suitable means (1), mixed models for repeated measurements are appropriate to address missing observations if it would not bias effects in favor of the treatment under evaluation. Anyhow, sensitivity analysis should be considered to explore robustness of the main estimator to deviations from its underlying assumptions and limitations in the data e.g., MNAR, MAR assumptions (28).

**Discussion and recommendation**

Missing values in clinical trials bear the risk of informing biased estimates. Despite numerous strategies to avoid missing values as early as during planning and conduct of the study, missing values will remain inevitable – with consequences for the assessment of added benefit accordingly.

A key aspect is the understanding of the missingness mechanism (i.e. MCAR, MAR of MNAR) and to which extent it can be considered non-informative. In general, statistical approaches assume the pattern of missingness to be at random (MAR) which rarely can be tested, but often can be justified by expert knowledge.

Strategies to handle missing values should be considered in the context of the estimand of interest. IQWiG is focused on the treatment policy approach. Other approaches may be considered only supportive but unlikely providing sufficient evidence for HTA decision making.

EUnetHTA recommends replacing missing data "*with a value derived from hypotheses about the HRQoL of patients with missing data*"(54). Single items missing in a HRQoL questionnaire will impact the scoring and thus imputation or certain algorithms should be used accordingly as long as number of missing data items is limited. Imputation examples are provided accordingly. When available, scoring manuals of the HRQoL instrument should be considered. EUnetHTA guidance also acknowledged several approaches that can be used to adjust for informative drop-out such as generalized linear mixed models and conditional linear models.

In practice, pre-planned single methods to deal with missingness may not be sufficient alone in order to meet expectations for HTA. Approaches like replacing missing values with single values for instance the subject's mean (or the mean of the subject's arm), the baseline or last observation carried forward (LOCF), or the worst case may require further exploration, and should not be used as the primary approach for missing data (52). Accordingly, handling of dropouts for longitudinal studies handling by MMRM approaches is considered favorable compared to LOCF (55). Application of various methods e.g., such as maximum likelihood based models (e.g., MMRM), multiple imputation, Bayesian, and weighted approaches could help to adequately address the uncertainty associated with the impact of missing values in the HTA context, in particular if the amount of missingness is high.

Regarding MMRM analysis, it seems reasonable in certain cases to consider a minimum amount of observations of a specific time period, for instance to ensure convergence of the model, although analysis with all observations is preferred.

The German benefit assessment will primarily evaluate patient relevant data when sufficiently complete. Thus, missing PRO data could be critical. The following recommendations may be helpful to increase acceptance of PRO for HA and HTA decision making in context of missing data.

- Every effort should be undertaken by study pharmaceutical companies to fulfil all the requirements of the protocol concerning the collection and management of data to limit missingness; this may include also routine monitoring of missing PRO data during study conduct to ensure that the PRO data are effectively collected according to the study protocol.
- Missing data need to be described in sufficient detail (reasons for discontinuation, frequency, and patient characteristics per group).

- Strategies for missing data analyses should primarily follow the treatment policy intercurrent event strategy for estimands.
- Supportive analyses are recommended to strengthen assumptions used for analysis e.g., to verify or visualize missing at random assumptions
    - Sensitivity analysis to explore the robustness of assumptions.
    - Complex replacement strategies require detailed operationalization in the dossier; thus, analysis plans should sufficiently describe the methods.
- Missing data exploration may be required for all PRO endpoints/concepts. For lung cancer studies typically cough, dyspnea and chest pain are key HRQoL concepts. However, complementary concepts to the specific questionnaire should be also investigated for the benefit assessment for the sake of completeness.
- Algorithms as laid down in the PRO scoring manuals shall be adopted to provide directions how to handle single-item missing values.
- Analysis planning of PRO data should carefully consider missingness.
- Scientific advice meetings should consider also to discuss appropriate approaches to deal with missing PRO data.
- Replacement strategies should consider the direction of bias due to missingness in the context of the underling objective, e.g., when PRO are used in context of tolerability and/or effectiveness.

# 9. PRO after treatment discontinuation

*Authors:     Sarah Böhme, Christoph Gerlinger, Friedhelm Leverkus, Michael Schlichting, Kati Sternberg*

**Background & rationale**

The National Research Council (NRC) recommended *"Trial pharmaceutical companies should continue to collect information on key outcomes on participants who discontinue their protocol-specified intervention in the course of the trial, except in those cases for which a compelling cost benefit analysis argues otherwise, and this information should be recorded and used in the analysis."*(56)

In general, PRO data collection period is often associated with the data collection period of key clinical endpoints e.g., tumor assessments in oncology studies, assessments of tolerability and safety as they aim to complement clinical efficacy and safety analyses and contextualize accordingly. Intercurrent events such as treatment discontinuation, disease progression, administration of rescue medication, or treatment switching can impact PRO analyses and interpretation.

The current EMA, *"Appendix 2 (6) to the Guideline on the evaluation of anticancer medicinal products in man – The use of patient-reported outcome (PRO) measures in oncology studies"*, recommends:

- PRO data collection over the clinically most important periods,
- duration of assessment should be limited to a time period that is both feasible and interpretable.

The assessment schedule should be terminated at a point when the results would no longer be interpretable either due to low compliance. They underline that the *"continued assessment post-progression and during next-line therapy may also be informative"* particularly in the palliative or maintenance setting *"when therapeutic claims (section 5.1 of the SmPC) are intended."* CHMP and also HTA bodies like IQWiG emphasize the issue of informative missing in case of varying follow-up times across treatment groups that could impact the conclusiveness of PRO analyses.

**Requirements by IQWiG and G-BA**

HTA agencies more and more request long-term, post-treatment PRO data. For instance, G-BA requests efficacy data (57) for new cancer drugs, including assessments of mortality, morbidity, and HRQoL, with specific recommendations to collect post progression HRQoL data. The main interest is on estimating the effect of a treatment strategy that includes the investigational treatment as compared to a treatment strategy that includes the adequate comparator. IQWiG requires a treatment policy approach irrespective of the underlying intercurrent event (e.g., such as treatment discontinuation due to toxicity or lack of efficacy (e.g., disease progression), switching to subsequent therapies etc. (see Figure 3) in order to minimize bias potentially introduced by varying follow-up times between treatment groups (2). Long-term follow-up data are suggested to be collected for all patients in oncology studies until drop out or death (2).

Figure 3: Illustrative Patient Treatment Journeys

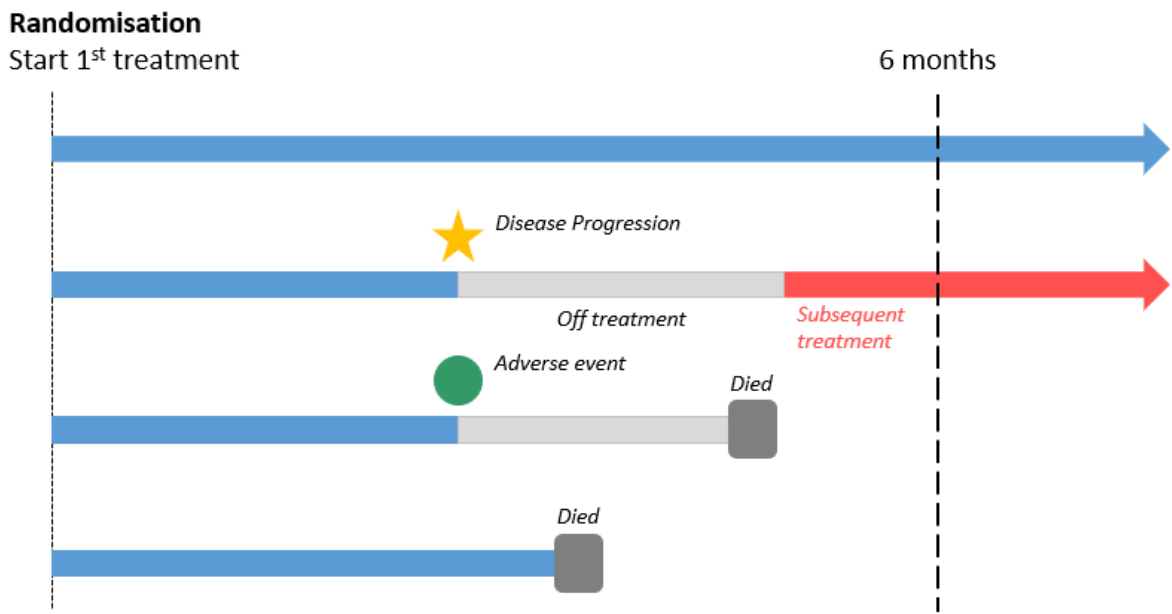

Note: used unchanged from (58) under a Creative Commons Attribution 4.0 International License (http://creativecommons.org/licenses/by/4.0/)

**Points to consider**

Feasibility of collecting long-term PRO data is associated with several challenges (59) and could also introduce systematic bias due to:

- Patient burden in particular in palliative settings impacting responsiveness and completion.
- Complexity, administrative burden to retrieve PRO data once the patient is off treatment.
- Protocol deviations: study protocols are less restrictive after the investigational period has been completed e.g., in terms of administering the PRO in right sequence, missingness of PRO administration, changing PRO administration mode.
- Treatment crossover / treatment switching: patients in the control group may crossover to the experimental treatment arm, different patients may receive different subsequent therapies based on their individual characteristics.

The estimation of treatment effects is complicated when patient attrition, loss-to follow-up, protocol deviations, non-compliance is apparent in long-term follow-up PRO investigations. Treatment switching / crossover may further confound estimation of the effect of the intervention. What is to be estimated (estimand) should be contextualized accordingly. The clinical utility might not be well understood in such situations and thus the question how such long-term PRO data could complement more objective endpoints remains unclear and overall the added value considering the costs may be challenging.

The current CHMP guidance on PRO data collection serve as the gold standard for study planning and thus will form the basis for HTA evaluations.

**Discussion and Methodological Recommendation**

Complete data capture from all patients is essential to limit bias and ensure generalizability of the analysis for the population of interest. Following the treatment policy strategy for intercurrent events as required by IQWiG, statistical analyses of all PRO measurements including those obtained after treatment discontinuation is essential for the benefit assessment in Germany. A while-on-treatment policy approach is frequently challenged and thus might be complemented by other intercurrent event strategies e.g., while-not-treated with subsequent anti-cancer therapy, to limit the risk of downgrading the added benefit rating.

In case long-term follow-up data are required that are not captured in the study, modelling or imputation approaches (hypothetical estimand) may be considered (see also missing data Section 8) but require sufficient justification and clarification of bias potential.

Joint HA and HTA scientific advice meetings may provide a unique opportunity to clarify and confirm relevant PRO research questions, if/how alternative ways to explore long-term PRO data could reasonably support the benefit-risk and the relative effectiveness assessments. Lundy et al (59) provided some interesting alternatives to collect long-term PRO data such as to collect long-term follow-up data from a subset of patients, consider different options for assessment venue (e.g., passive follow-up via registries, active follow-up via cohort extension), employ modes of PRO administration that can be used at the participant's home. However, IQWiG may not likely to accept them.

# 10. Discussion and Conclusion:

PRO are increasingly utilized and recognized as essential data towards patient focused drug development. Regulatory and HTA bodies consider patient perspectives in their decision making, to inform benefit-risk and also relative effectiveness assessments. In 2020, key guidances have been issued by various stakeholders that will impact PRO evaluations respectively:

- ICH E9 (R1) addendum on estimands and sensitivity analysis in clinical trials to the guideline on statistical principles for clinical trials by CHMP; EMA/CHMP/ICH/436221/2017 Committee for Medicinal Products for Human Use, 17 February 2020
- International standards for the analysis of quality-of-life and patient-reported outcome endpoints in cancer randomized controlled trials: recommendations of the SISAQOL Consortium by EORTC
- IQWiG General Methods Version 6, by November 5$^{th}$, 2020

SISAQOL has provided taxonomy of PRO research objectives for cancer trials and provide recommendations for endpoints and statistical analysis methods considering a broad representation of stakeholders. The estimand framework aims to improve planning, design, analysis and interpretation of clinical trials. There are also regulatory guidance documents in development that structure the way evidence is synthesized using patient perspectives.

We acknowledge the different perspectives of the new guidance documents. Local methodological guidances could raise new challenges and risks for local reimbursement that could be best mitigated in close collaboration with local experts and statisticians.

For instance, different thresholds are considered to indicate a relevant response threshold for PRO, either in terms of how such thresholds are determined and how they are used for decision making. The IQWiG General Methods Paper suggests a 15% change of the scale range as a universal relevant individual threshold, whereas regulators recommend an anchor-based approach. A variety of analyses need to be prepared with potential different outcomes to address different research questions accordingly. Currently G-BA recommends submitting analyses based on the 15% criteria and also established thresholds for the added benefit assessment. However, for norm-based scale scores such as SF36, it appears unclear if a practical or a theoretical scale range is appropriate. The scale range could vary, based on the norm population investigated or it could be derived theoretically in case of no missing data. According to the transcription of the oral G-BA hearing of Secukinumab (31), G-BA expressed the need for further internal discussions to better understand the practical implications of IQWiG's new approach. In summary, the volume of PRO analyses using a response threshold will increase as will the likelihood of discordant conclusions on the relevant estimands between G-BA and regulatory bodies, SMPC and treatment guidance such as S3 guidance based on different response thresholds.

There are also specific recommendations to collect long-term follow-up data for PRO. It is suggested to collect PRO data for all patients in oncology studies until drop out or death post progression. IQWiG requires to estimate the effect considering a treatment policy approach irrespective of the underlying intercurrent event to minimize bias potentially introduced by varying follow-up times between treatment groups. However, the underlying research question and estimand might be impacted by the feasibility of collecting long-term PRO data considering challenges such as patient attrition, protocol deviations, treatment switching etc. How such long-term PRO data could complement more objective endpoints e.g., anti-tumor activity, that is being collected until treatment discontinuation, remains still unclear.

We highlighted also the challenges associated to obtain an added benefit claim for PRO including but not limited to the different requirements on the PRO instrument and methodology. Anyway, requirements by IQWiG and G-BA must be considered for the AMNOG dossier. Currently, a notable amount of necessary PRO analyses for the benefit assessment are not usually specified in the Statistical Analysis Plan that will be carried out for the Clinical Study Report. It is thus recommended to draft an HTA specific analysis plan for the benefit assessment covering IQWiG´s methodological requirements and considering G-BA guidances. Although HTA data analysis will be performed retrospectively, completeness of PRO data and analyses including all PRO endpoints, even when initially planned as tertiary endpoints, is key to succeed for the benefit assessment of drugs. Study planning, conduct and analyses should take into account HTA specific PRO research questions to facilitate evaluations required for reimbursement purposes.

Pharmaceutical companies should early reflect on the validity of instruments and associated thresholds that determine PRO response, consider PRO requirements that are relevant for HTA decision making when designing a study. A risk assessment may help to identify gaps, initiate further data generation plans to support evidence synthesis for PRO which will include but is not limited to comprehensive PRO analysis planning. Early scientific advice meetings with G-BA are highly recommended to clarify and confirm if/how such methodological challenges could be resolved in terms of study design, conduct and also analyses to adequately reflect the patient perspective in decision making.

# Literature reference

1. Gemeinsamer Bundesausschuss. Verfahrensordnung des Gemeinsamens Bundesausschusses; 2021.
2. Gesundheitswesen IfQuWi. General Methods; Version 6.0 of 5 November 2020. 2020.
3. Gemeinsamer Bundesausschuss. Modul 4 – Medizinischer Nutzen und medizinischer Zusatznutzen, Patientengruppen mit therapeutisch bedeutsamem Zusatznutzen Template; [Template]. 2013.
4. Böhme S, Genet A, Gillhaus J, Kürschner N, Leverkus F, Schiffner-Rohe J, et al. Benefit assessment in Germany: requirements & challenges presented for 6 topics. In: GmbH PD, editor. 2017.
5. Dane A, Spencer A, Rosenkranz G, Lipkovich I, Parke T. Subgroup analysis and interpretation for phase 3 confirmatory trials: White paper of the EFSPI/PSI working group on subgroup analysis. Pharmaceutical Statistics. 2018;18(2):126-39.
6. European Medicines Agency. Appendix 2 to the guideline on the evaluation of anticancer medicinal products in man2014 https://www.ema.europa.eu/en/documents/other/appendix-2-guideline-evaluation-anticancer-medicinal-products-man_en.pdf.
7. Deshpande PR, Rajan S, Sudeepthi BL, Abdul Nazir CP. Patient-reported outcomes: A new era in clinical research. Perspectives in Clinical Research. 2011;2(4):137-44.
8. U.S. Department of Health and Human Services FDA Center for Drug Evaluation and Research, U.S. Department of Health and Human Services FDA Center for Biologics Evaluation and Research, U.S. Department of Health and Human Services FDA Center for Devices and Radiological Health. Guidance for industry: patient-reported outcome measures: use in medical product development to support labeling claims: draft guidance. Health and Quality of Life Outcomes. 2006;4(79).
9. Rotenstein LS, Huckman RS, Wagle NW. Making Patients and Doctors Happier — The Potential of Patient-Reported Outcomes. The New England Journal of Medicine. 2017;377(14):1309-12.
10. U.S. Department of Health and Human Services FDA, Center for Drug Evaluation and Research. Clinical Outcome Assessment (COA) Compendium2019 https://www.fda.gov/media/130138/download.
11. U.S. Department of Health and Human Services FDA. Clinical Outcome Assessment (COA): Frequently Asked Questions. What is a COA and what are the different types? https://www.fda.gov/about-fda/clinical-outcome-assessment-coa-frequently-asked-questions#COADefinition 2020.
12. European Medicines Agency. Reflection Paper on the use of patient reported outcome (PRO) measures in oncology studies2014 https://www.ema.europa.eu/en/documents/scientific-guideline/draft-reflection-paper-use-patient-reported-outcome-pro-measures-oncology-studies_en.pdf.
13. Brettschneider C, Lühmann D, Raspe H. Der Stellenwert von Patient Reported Outcomes (PRO) im Kontext von Health Technology Assessment (HTA) 2011.
14. U.S. Department of Health and Human Services FDA, Center for Drug Evaluation and Research, Center for Biologics Evaluation and Research, Center for Devices and Radiological Health. Guidance for Industry. Patient-Reported Outcome Measures: Use in Medical Product Development to Support Labeling Claims 2009 https://www.fda.gov/media/77832/download.
15. U.S. Department of Health and Human Services FDA Center for Biologics Evaluation and Research, Center for Devices and Radiological Health, Research CfBEa. Principles for Selecting, Developing, Modifying, and Adapting Patient-Reported Outcome Instruments for Use in Medical Device Evaluation. Draft Guidance for Industry and Food and Drug

Administration Staff, And Other Stakeholders2020 https://www.fda.gov/media/141565/download.

16. U.S. Food & Drug Administration. FDA Patient-Focused Drug Development Guidance Series for Enhancing the Incorporation of the Patient's Voice in Medical Product Development and Regulatory Decision Making 2020. URL: https://www.fda.gov/drugs/development-approval-process-drugs/fda-patient-focused-drug-development-guidance-series-enhancing-incorporation-patients-voice-medical.
17. European Medicines Agency. Reflection Paper on the Regulatory Guidance for the Use of Health-Related Quality of Life (HRQL) Measures in the Evaluatio of Medicinal Products 2005.
18. International Society for Pharmacoeconomics and Outcomes Research. Measurement Comparability Between Modes of Administration of PROMs; Task Force. URL: https://www.ispor.org/member-groups/task-forces/measurement-comparability-between-modes-of-administration-of-proms.
19. International Society for Pharmacoeconomics and Outcomes Research. Patient Reported Outcomes: Analysis and Interpretation. URL: https://www.ispor.org/conferences-education/education-training/virtual/distance-learning/patient-reported-outcomes-analysis-and-interpretation.
20. Rothman M, Burke L, Erickson P, Kline Leidy N. Use of Existing Patient-Reported Outcome (PRO) Instruments and Their Modification: The ISPOR Good Research Practices for Evaluating and Documenting Content Validity for the Use of Existing Instruments and Their Modification PRO Task Force Report. Value in Health. 2009;12(8):1075-83.
21. International Society for Quality of Life Research. Implementing Patient-Reported Outcome Measures in Clinical Practice: A Companion Guide to the ISOQOL User's Guide2018 https://www.isoqol.org/wp-content/uploads/2019/09/ISOQOL-Companion-Guide-FINAL.pdf.
22. Coens C, Pe M, Dueck A, Sloan J, Basch E. International standards for the analysis of quality-of-life and patient-reported outcome endpoints in cancer randomised controlled trials: recommendations of the SISAQOL Consortium. The Lancet Oncology. 2020;21(2):e83-e96.
23. Mokkink LB, Terwee CB, Patrick DL, Alonso J, Stratford PW, Knol DL, et al. The COSMIN checklist for assessing the methodological quality of studies on measurement properties of health status measurement instruments: an international Delphi study. Quality of Life Research. 2010;19:539-49.
24. Prinsen CAC, Mokkink LB, Bouter LM, Alonso J, Patrick DL, De Vet HCW, et al. COSMIN guideline for systematic reviews of patient-reported outcome measures Quality of Life Research. 2018;27(5):1147-57.
25. Calvert M, Blazeby J, Altman DG, Revicki DA, Moher D. Reporting of Patient-Reported Outcomes in Randomized Trials. The CONSORT PRO Extension. Journal of the American Medical Association. 2013;309(8):814-22.
26. Calvert M, Kyte D, Mercieca-Bebber R. Guidelines for Inclusion of Patient-Reported Outcomes in Clinical Trial Protocols. The SPIRIT-PRO Extension. Journal of the American Medical Association. 2018;319(5):483-94.
27. Williamson P, Gargon E, Bagley H, Blazeby J, Clarke M. Core Outcome Measures in Effectiveness Trials [Zugriff: https://www.cometinitiative.org/].
28. European Medicines Agency. ICH E9 (R1) addendum on estimands and sensitivity analysis in clinical trials to the guideline on statistical principles for clinical trials 2020.
29. Gemeinsamer Bundesausschuss. Beschluss des Gemeinsamen Bundesausschusses über eine Änderung der Arzneimittel-Richtlinie (AM-RL): Anlage XII - Beschlüsse über die Nutzenbewertung von Arzneimitteln mit neuen Wirkstoffen nach § 35a SGB V – Crizotinib. 2016.
30. Gemeinsamer Bundesausschuss. Beschluss des Gemeinsamen Bundesausschusses über eine Änderung der Arzneimittel-Richtlinie (AM-RL): Anlage XII - Beschlüsse über die Nutzenbewertung von Arzneimitteln mit neuen Wirkstoffen nach § 35a SGB V – Eribulin (neues Anwendungsgebiet). 2016.

31. Gemeinsamer Bundesausschuss. Beschluss des Gemeinsamen Bundesausschusses über eine Änderung der Arzneimittel-Richtlinie (AM-RL): Anlage XII – Nutzenbewertung von Arzneimitteln mit neuen Wirkstoffen nach § 35a SGB V Secukinumab (Neubewertung aufgrund neuer Wissenschaftlicher Erkenntnisse (PsoriasisArthritis)). 2021.
32. Leverkus F. Methodik für die Auswertung von Lebensqualitätsdaten im Rahmen der Nutzenbewertung von Arzneimitteln. Zeitschrift für Evidenz, Fortbildung und Qualität im Gesundheitswesen. 2014;108:111-9.
33. Gerlinger C, Bamber L, Leverkus F, Schwenke C, Haberland C. Comparing the EQ-5D-5L utility index based on value sets of different countries: impact on the interpretation of clinical study results. BMC Research Notes. 2019;12(18).
34. Duru G, Auray JP, Béresniak A, Lamure M, Paine A. Limitations of the methods used for calculating quality-adjusted life-year values. Pharmacoeconomics. 2002;20(7):463-73.
35. Klakow-Franck R. Die Bedeutung von Lebensqualität für die Arbeit des Gemeinsamen Bundesausschusses. Zeitschrift für Evidenz, Fortbildung und Qualität im Gesundheitswesen. 2014;108(2-3):151-6.
36. Kohlmann T. Messung von Lebensqualität: So einfach wie möglich, so differenziert wie nötig. Zeitschrift für Evidenz, Fortbildung und Qualität im Gesundheitswesen. 2014;108:104-10.
37. Bullinger M, Hasford J. Evaluating quality-of-life measures for clinical trials in Germany. Controlled Clinical Trials. 1991;12:915-105S.
38. Osaba D, Zee B, Pater J, Warr D, Kaizer L, Latreille J. Psychometric properties and responsiveness of the EORTC quality of Life Questionnaire (QLQ-C30) in patients with breast, ovarian and lung cancer. Quality of Life Research. 1994;3:353-64.
39. Cella D, Tulsky D, Gray G, Sarafian B, Linn E, Bonomi A. The Functional Assessment of Cancer Therapy scale: development and validation of the general measure. Journal of Clinical Oncology. 1993;11(3):570-9.
40. McHorney C, Ware J, Raczek A. The MOS 36-Item Short-Form Health Survey (SF-36): II. Psychometric and Clinical Tests of Validity in measuring Physical and Mental Health Constructs. Medical Care. 1993;1993(31):3.
41. McHorney C, Ware J, Lu J, Sherbourne C. The MOS 36-Item Short-Form Health Survey (SF-36): III. Tests of Data Quality, Scaling Assumptions, and Reliability Across Diverse Patient Groups. Medical Care. 1994;1994(32):1.
42. Schünemann HJ, Guyatt GH. Commentary--goodbye M(C)ID! Hello MID, where do you come from? Health Services Research. 2005;40(2):593-7.
43. Jaeschke R, Singer J, Guyatt GH. Ascertaining the Minimal Clinically Important Difference Controlled Clinical Trials. 1989;10:407-15.
44. Mouelhi Y, Jouve E, Castelli C, Gentile S. How is the minimal clinically important difference established in health-related quality of life instruments? Review of anchors and methods. Health and Quality of Life Outcomes. 2020;18(136).
45. McGlothlin AE, Lewis RJ. Minimal Clinically Important Difference. Defining What Really Matters to Patients. Journal of the American Medical Association. 2014;312(13):1342-43.
46. Devji T, Carrasco-Labra A, Qasim A, Phillips M, Johnston BC, Devasenapathy N. Evaluating the credibility of anchor based estimates of minimal important differences for patient reported outcomes: instrument development and reliability study. BMJ. 2020;369.
47. Gemeinsamer Bundesausschuss. Beschluss des Gemeinsamen Bundesausschusses über eine Änderung der Arzneimittel-Richtlinie (AM-RL): Anlage XII – Nutzenbewertung von Arzneimitteln mit neuen Wirkstoffen nach § 35a SGB V Indacaterolacetat/Glycopyrroniumbromid/Mometasonfuroat (Asthma). 2021.
48. Gemeinsamer Bundesausschuss. Antworten auf häufig gestellte Fragen zum Verfahren der Nutzenbewertung. URL: https://www.g-ba.de/themen/arzneimittel/arzneimittel-richtlinie-anlagen/nutzenbewertung-35a/faqs/.
49. Gemeinsamer Bundesausschuss. Beschluss des Gemeinsamen Bundesausschusses über eine Änderung der Arzneimittel-Richtlinie (AM-RL): Anlage XII – Nutzenbewertung von

50. Skipka G, Wieseler B, Kaiser T, Thomas S, Bender R, Windeler J. Methodological approach to determine minor, considerable, and major treatment effects in the early benefit assessment of new drugs. Biometrical Journal. 2016;58(1):43-58.
51. European Medicines Agency. ICH Topic E 9. Statistical Principles for Clinical Trials. 1998.
52. European Medicines Agency. Guideline on Missing Data in Confirmatory Clinical Trials. 2010.
53. Little RJA, Rubin DB. Statistical analysis with missing data: Wiley; 2019.
54. European Network for Health Technology Assessment. Guideline: Endpoints used for Relative Effectiveness Assessment: Health-Related Quality of Life and Utility Measures. 2015.
55. Lane P. Handling drop-out in longitudinal clinical trials: a comparison of the LOCF and MMRM approaches. Pharmaceutical Statistics. 2008;7:93-106.
56. National Research Council (US) Panel on Handling Missing Data in Clinical Trials. The Prevention and Treatment of Missing Data in Clinical Trials. Washington (DC): National Academies Press (US); 2010.
57. Gemeinsamer Bundesausschuss. The benefit assessment of medicinal products in accordance with the German Social Code, Book Five (SGB V), section 35a. URL: https://www.g-ba.de/english/benefitassessment/.
58. Lawrance R, Degtyarev E, Griffiths P, Trask P, Lau H, D'Alessio D, et al. What is an estimand & how does it relate to quantifying the effect of treatment on patient-reported quality of life outcomes in clinical trials? Journal of Patient-Reported Outcomes. 2020;4(68).
59. Lundy JJ, Coon CD, Fu A-C, Pawar V. Collection of Post-treatment PRO Data in Oncology Clinical Trials. Therapeutic Innovation & Regulatory Science. 2021;55:111-7.

Arzneimitteln mit neuen Wirkstoffen nach § 35a SGB V Nintedanib (neues Anwendungsgebiet: interstitielle Lungenerkrankung mit systemischer Sklerose (SSc-ILD)) 2021.